\documentstyle[11pt,pazh2,psfig]{article}
\newcommand{\km}{\,\mbox{km}\,\mbox{s}^{-1}}

\begin{document}
\title{Interstellar Medium Surrounding the WO Star in the Galaxy IC 1613:
New Optical and Radio Observations\thanks{Astronomy  Reports,  Vol. 45 ,
No~6, 2001,  Translated  from  Astronomicheskii
Zhurnal, Vol. 78, No. 6. Translated by D. Gabuzda}}

\author{T.A.Lozinskaya \inst{a} \and A.V.Moisseev\inst{b} \and
V.L. Afanasiev\inst{b} \and
E. Wilcots\inst{c} \and  W.M.Goss \inst{d}}

\institute{
  Sternberg Astronomical Institute, Universitetskii pr. 13,
Moscow, 119899 Russia \and
 Special Astrophysical Observatory, Nizhnii Arkhyz,
Karachaevo-Cherkessia, 369167 Russia
\and
  Department of Astronomy, University of Wisconsin -
Madison, 475 N. Charter St., Madison, WI 53706, USA
\and
 National Radio Astronomy Observatory, New Mexico Array
Operations Center, P.O. Box O, 1003 Lopezville Road, Socorro,
NM 87801
}

\maketitle
\begin{abstract}

{\small
Observations of the nebula S3 associated with the WO
star in the galaxy IC 1613 and of an extended region surrounding S3 are
reported. The star and bright core of the nebula were observed
with a multipupil fiber spectrograph mounted on the 6-m telescope
of the Special Astrophysical Observatory of the Russian Academy of
Sciences. Images in the principle spectral lines and integrated spectra
of the star and three compact clumps were obtained, and the
radial-velocity field constructed.
An extended region of the galaxy was observed with the Very Large Array
at 21 cm. A giant ring or H~I shell enclosing a large fraction of the
stellar population in IC 1613 was discovered. The WO star and
associated bipolar nebula, which we discovered earlier, lies at the
inner edge of the H~I ring. A local H~I deficiency and two arc-like
H~I ridges were also detected for the first time, and probably represent
the neutral component of the bipolar shell surrounding the WO star.
The two arc-like ridges may also have been produced by the collective
stellar wind (and supernova explosions?) in OB association No. 9 from
the list of Hodge. A scenario for the formation of the extended bipolar
feature is discussed, based on the new data.}
\end{abstract}

\small
\section{Introduction}
     Only  six  WR stars of more than five hundred such objects in
the  Local  Group  of  galaxies belong to the rare WO class, which
represents  a very brief final evolutionary stage of massive stars
approaching  the naked CO-core stage (see Barlow \& Hummer (1982),
Kingsburg  et  al.  (1995)  and  references therein). WO stars are
characterized  by  powerful stellar winds with terminal velocities
reaching   $\sim   5000\km$   (  Barlow \& Hummer,1982; Torres  et
al., 1986;  Dopita  et  al.,  1990; Potcaro et al., 1992 ) and their
effective  temperatures  can be as high as $10^5$~K (Dopita et al,
1990;  Maeder  \&  Meynet,  1989; Melnik \& Heydari-Malayeri, 1991;
Polcaro  et  al., 1991).  One  of these six stars was discovered by
D'Odorico  \&  Rosa  (1982)  (see  also  Davidson  \& Kinman 1982;
Armandroff  \&  Massey,  1985)  in  the  dwarf irregular galaxy IC
1613.  The  star  was  identified  by  the  specific wide bands it
produces  in  the  spectrum  of  the  core  of the bright emission
nebula   S3   from   the   list  of  Sandage  (1971)  --  a  giant
$29''\times9''   (93\times30$~pc)  H~II  region.  Throughout  this
paper,  we  adopt  the distance 660 kpc for IC 1613, as derived by
Saha  et  al.  (1992). The central region of S3 shows bright He~II
4686\AA~  emission  (D'Odorico \& Rosa, 1982; Smith (1995), Garnett
et  al.,  1991).  A detailed spectral analysis of the nebula and a
discussion  of  its  chemical composition are presented by Garnett
et  al.(1991)  and  Kingsburgh \& Barlow (1995) -- hereafter KB95.
The  nebula  S3 was found to be a source of thermal radio emission
(Goss   \&  Lozinskaya,  1995);  the  full  size  at  half-maximum
intensity    of   the   radio   source   ($19^{\prime\prime}\times
14^{\prime\prime}$)  corresponds to a bright region of the optical
nebula  visible  on  deep  H$\alpha$  images taken by Hodge et al.
(1990) and Hunter et al. (1993).

%\bigskip
\begin{table*}
\caption{ Log of MPFS observations.}
\begin{center}
\begin{tabular}{|c|c|c|c|c|c|c|}
\hline
Date  & Size of    & Field of view &Wavelenth &$\delta\lambda$ &Exposure& Seeing\\
      & micropupil &               &interval&                & s  & \\
\hline
29.XI.99 &$1^{\prime\prime}$   &$16^{\prime\prime}\times15^{\prime\prime}$& $4450-7050$\AA& 8\AA          & 7200 &
$2.^{\prime\prime}7$ \\
30.XI.99 &$0.^{\prime\prime}75$&$12^{\prime\prime}\times11.3^{\prime\prime} $& $4450-7050$\AA& 8\AA          & 7200 &
$1.^{\prime\prime}8$ \\
\hline
\end{tabular}
\end{center}
\end{table*}

Our observations made with a scanning Fabry-Perot interferometer
mounted on the 6-m telescope of the Special Astrophysical Observatory
(SAO) and narrow-band H$\alpha$ and [O~III] images taken with the 4-m
telescope of the Kitt Peak National Observatory resulted in the
discovery of a faint external bipolar feature (hereafter called the
bipolar   shell)   extending  far  beyond  the  bright  nebula  S3
(Lozinskaya, 1997; Afanasiev et al., 2000 -- Paper~I).
The sizes of the two external components--the southeastern and
northwestern shells--are $35^{\prime\prime}\times 24^{\prime\prime}
(112\times77$~pc) and $\simeq 60^{\prime\prime}\times 70^{\prime\prime}
(190\times220$~pc), respectively. Radial-velocity measurements at the
maximum and the H$\alpha$ line halfwidths provide
evidence for expansion of the bipolar shell, with lower limits to the
expansion velocities of the southeastern and northwestern shells
being 50 and at least $70\km$, respectively.

In  Paper~I we  proposed a scenario for the formation
of the unique bipolar feature by the stellar wind of the WO star,
which is embedded in a dense layer at the boundary of a supercavity
in the galaxy's H~I distribution. We identified this supercavity by
analyzing the H~I column density distribution in IC 1613 published
by Lake and Skillman (1989). At that time, we hypothesized the presence
of a dense shell surrounding the supercavity and a physical association
with the WO star, but were not able to prove that this was the case.

In the current paper, we report the results of new observations of the
nebula associated with the WO star in the galaxy IC 1613. We observed
the bright core of the nebula S3 using the MPFS multipupil fiber
spectrograph mounted on the SAO 6-m telescope (Section 2). We obtained
integrated spectra of individual compact clumps in the central bright
region, and reconstructed the radial-velocity field from our H$\beta$
and [O~III] 5007\AA~ measurements. We also obtained observations of a
large sector of IC 1613 at 21 cm with the Very Large Array (VLA; Section
3). We derived a detailed image of the neutral-gas distribution, which
revealed a giant dense ring or H~I shell surrounding the supercavity,
and showed that the WO star does indeed reside in this dense ring and
is physically interacting with it. We also found a local H~I deficiency
and two extended features, testifying to the presence of a neutral gas
component associated with the ionized bipolar shell. In Section 4,
we discuss the results of the observations and analyze a possible
scenario for the formation of the extended bipolar feature.

\section{ Spectroscopic observations of the WO star
and nebula S3}

\subsection{ Observations and Data Reduction}

The spectroscopic observations were made on November 29 and 30, 1999
using  the new integral-field spectrograph MPFS (multipupil fiber
spectrograph) mounted at
the primary focus of the SAO 6-m telescope. A description of the
spectrograph can be found at the web-page\\
\verb*"http://www.sao.ru/~gafan/devices.htm"

Compared to the earlier version of the Afanasiev et al.(1990), the new spectrograph
has a larger field of view, wider spectral range, and higher quantum
efficiency. The spectrograph uses a $1024\times 1024$ CCD as a detector
and allows a set of spectra to be taken simultaneously from 240 spatial
elements (`micropupils') forming a $16\times 15$ array in the plane of
the sky. The angular size of an image element can be varied from 0\farcs5 to
$1.0^{\prime\prime}$. A spectrum of the sky background 4\farcm5
from the center of the field of view is taken simultaneously. Table 1 gives a
log of the MPFS observations.

These observations were reduced using IDL-based software developed
by the SAO Laboratory for Spectroscopy and Photometry of Extragalactic
Objects. The preliminary data reduction included bias-frame subtraction,
flat-field correction, cosmic-hits removal, extraction of individual
spectra from the CCD images, and wavelength calibration using the
spectrum of a He-Ne-Ar lamp. We then subtracted the night-sky spectrum
from the linearized spectra and used a number of spectrophotometric
standards observed at the same zenith angle as IC 1613 to convert the
observed fluxes to an absolute energy scale. We used the same standards
to derive the seeing estimates given in Table 1.

\begin{figure*}
\centerline{\psfig{figure=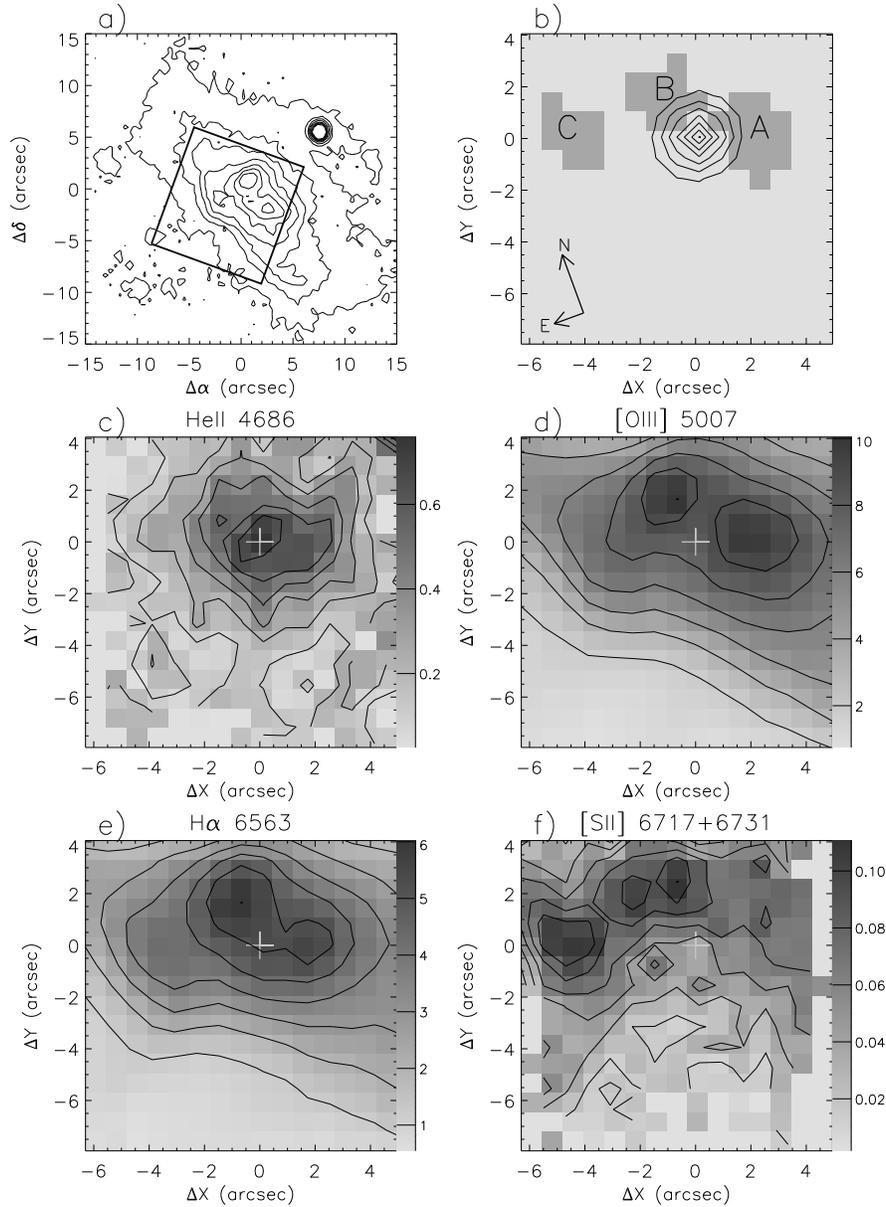,width=12 cm}}
\caption{
(a) [O~III] image of the nebula surrounding the WO star,
adopted from Paper~I. The frame shows the MPFS field of view for the
observations described in this paper. (b) MPFS observations.
Isophotes show the brightness distribution in the C~IV~$5801$\AA~ line
emitted by the WO star. Also indicated are the positions of emission
clumps. (c), (d), (e), and (f): Images and isophotes of the
nebula in the He~II~4686\AA, [O~III]~5007\AA, H$\alpha$, and
[S~II]~6717+6731\AA~emission lines, respectively. The cross
indicates the position of the WO star. Surface brightness is in units
of $10^{-15}$~erg~cm$^{-2}$~s$^{-1}$~arcsec$^{-2}$.
}
\end{figure*}

Analysis of the observations showed that the data obtained on
November 29 and 30, 1999 yielded consistent values for the main
parameters measured (line fluxes and velocities). Therefore, we will
consider further only the November 30, 1999 data, since they had higher
angular resolution.

Figure 1a shows an [O~III] line image of the central region of the
nebula  S3  surrounding  the WO star from the data of Paper~I,  together
with the boundaries of the field of view of our new MPFS observations.

\subsection{The WO star.}

We obtained the integrated spectrum of the WO star by co-adding 24
individual spectra, which exhibited the broad emission bands
characteristic of Wolf-Rayet stars.

During the flux measurements, we fitted the emission lines using
superpositions of two to six Gaussians; the resulting flux and
equivalent-width estimates are summarized in Table 2. Note that
the errors quoted in this table do not include uncertainty in the
continuum level, which could contribute systematic errors to the
equivalent widths. For comparison, this table also gives the fluxes
and equivalent widths of the corresponding lines from KB95.

We observed a narrower spectral interval than KB96;
it is clear from Table 2 that the two sets of observations do not
show significant discrepancies. The differences between the equivalent
widths are due primarily to the different methods used to determine the
continuum level (unlike KB95, we did not take
into account the nebular continuum component). Our flux estimates are
higher than those of KB95 for most of the lines. This is likely due
to the fact that our panoramic detector collected all the light from the
star, whereas a long slit whose width ($1^{\prime\prime}$) was comparable to or
smaller than the image size was used in KB95. To test this hypothesis,
we integrated the spectra of the star only over the lenses lying
within a $1^{\prime\prime}\times 5^{\prime\prime}$ rectangle aligned
along $PA = 55^{\circ}$, i.e., in the direction of the slit used in KB95.
The resulting WO-star line fluxes, indeed, proved to be a factor of 1.5--2
lower, on average, than the fluxes listed in Table 2. Moreover, as we
show below, the emission-line brightness peak in the nebula does not
coincide with the position of the star. This could have led to an offset
during the positioning of the narrow slit on the star in KB95, resulting
in further underestimation of the line fluxes.

We constructed images in the C~IV+He~II 4659\AA, O~V 5590\AA, and
C~IV 5801\AA~emission lines by integrating the spectra over the intervals
$4600-4725$, $5535-5653$, and $5735-5850$\AA, after subtraction of
narrow nebular line components blended with the WO-star lines. The
centroids of all the images agree to within $\pm0.05^{\prime\prime}$, and also
agree with the centroid of a continuum image made in the vicinity of
6100\AA. We estimate the FWHM of these emission-line images of the
WO star to be $1.85\pm0.05^{\prime\prime}$, $1.84\pm0.09^{\prime\prime}$,
and $1.67\pm0.03^{\prime\prime}$,
respectively, consistent with the seeing estimate of 1\farcm8
obtained for the standard star (Table 1). Note that, due to certain
specific features of the spectrograph, the images constructed from
the micropupil array are slightly elongated (1:1.15) along the Y axis
(Fig. 1b). Therefore, all our FWHM estimates are based on measurements
in the X direction.

The continuum image of the WO star was somewhat more extended
(FWHM = $2.2\pm 0.15^{\prime\prime}$) and exhibited a faint base. (This could
indicate the presence of a compact cluster made up of several stars,
but firm conclusions require further observations).

\subsection{Structure~and~emission~spectrum~of~S3}

We constructed an entire series of monochromatic images of the bright
core of the nebula in the He~II~4686\AA, [O~III]~5007\AA,
[S~II]~6717+6731\AA, and H$\alpha$ lines by obtaining Gaussian fits
to the emission lines. The results are shown in Figs. 1c--1f. All the
line profiles are fit well by single Gaussians, and do not show any
systematic deviations. Note, however, that the expected relative gas
velocities (up to $100\km$ in the bright core Paper~I) are appreciably
lower than the spectral resolution of our data ($\mbox{FWHM} =
350-450\km$).

\begin{figure*}
\psfig{figure=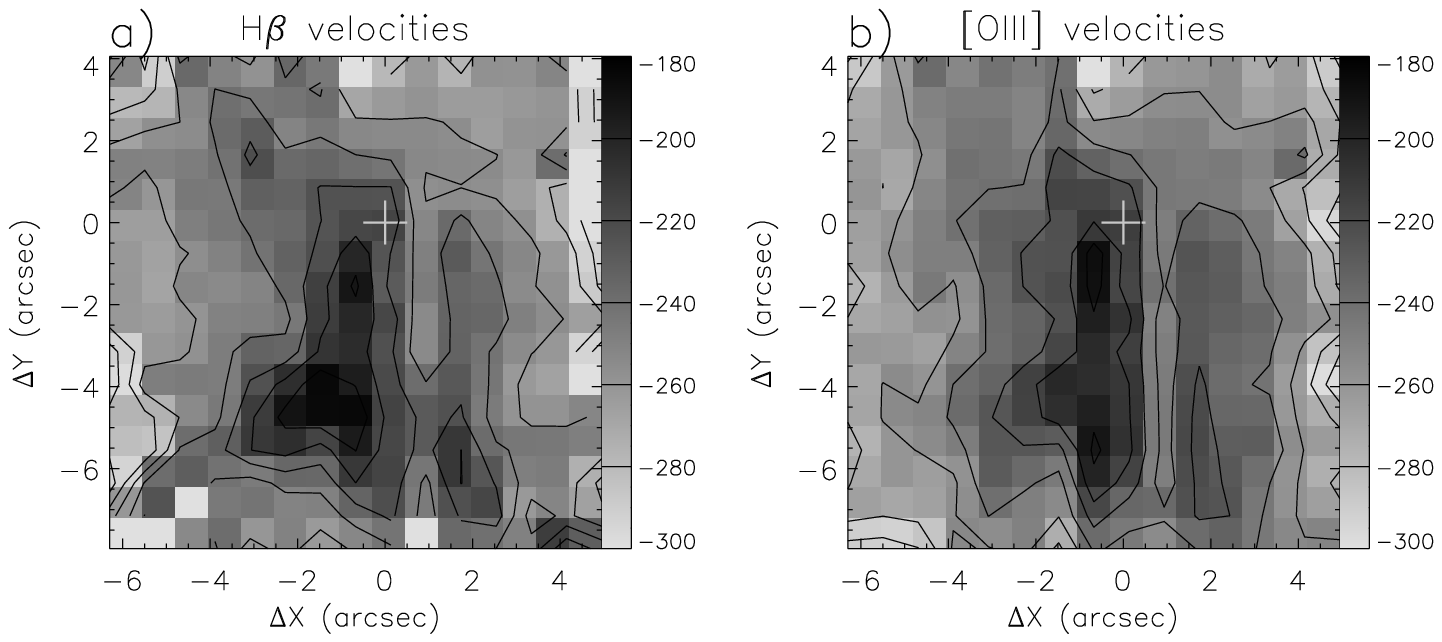,width=16 cm}
\caption{Ionized gas velocity field in the (a) H$\beta$ and the (b)
[O~III]~5007\AA~lines based on the MPFS observations. The cross
indicates the position of the WO star. The velocity scale is in $\km$.
}
\end{figure*}

It is evident from Fig. 1 that the morphology of the nebula is appreciably
different in different lines. For example, the He~II brightness peak
is coincident with the position of the WO star; the He~II brightness
decreases with distance from the star, and the full size of the emission
region is $10^{\prime\prime}\times 9^{\prime\prime}$. The situation is
quite different in the brighter [O~III] and H$\alpha$ lines, and also
in the [S~II] lines: the corresponding images exhibit several bright
clumps $2-3^{\prime\prime}$ from the star (Fig. 1). The [O~III] and
H$\alpha$ images show two well-defined
bright clumps north and northwest of the star (marked A and B,
respectively, in Fig. 1b), and a more extended clump northeast of the
star (clump C). This extended clump is especially prominent in the
[S~II] images, where clump A cannot be seen at all.

We discovered the clumpy structure of the S3 core in our earlier
study in Paper~I, based on uncalibrated [O~III] and H$\alpha$ images
of the region. However, having no emission spectra for the clumps,
we mistook the WO star for a clump (clump No. 2 in Paper~I). Clumps A
and B from our new observations correspond to clumps 3 and 1,
respectively, from Paper~I; clump C, which can be confidently identified
only in the [S~II] lines, was not discovered earlier in the [O~III] and
H$\alpha$ images.

We obtained the integrated spectra of clumps A, B, and C; the
positions of the image elements whose spectra were co-added in each
clump are shown in Fig. 1b. Table 3 gives the relative line intensities
for the three clumps, corrected for selective extinction. We assumed
c(H$\beta$) = 0.1 and E(B--V)=0.069, in accordance with KB95.
For comparison, Table 3 also presents the data of KB95 for the
integrated spectrum of the central region of S3.

\subsection{Ionized~Gas~Velocities in the Core~of~S3}

We used the resulting spectra to construct the radial-velocity field
for the central part of the nebula S3. We determined the velocity
fields from measurements made in the main spectral lines from Gaussian
fits to the corresponding line profiles. Analysis of the night-sky spectrum
showed that, after linearization of the spectra, there remained
systematic velocity variations in the night-sky lines across the field
of $\pm 0.1-0.15$~channels, i.e., about $15\km$. We corrected our
emission-line velocity fields for this systematic offset; Fig. 2 shows
the results.

We used the resulting velocity fields to estimate the mean
velocities in various lines for clumps A, B, and C and for the
nebula region projected onto the WO star. In this last case, we used
the same set of lenses used when co-adding the spectrum of the star
(see Section 2.2.). The results are summarized in Table 4, where the
quoted errors correspond to the velocity dispersion within individual
spatial elements.

\begin{table*}
\caption{
Fluxes and equivalent widths of emission lines in the
spectrum of the WO star.
}
\begin{center}
\begin{tabular}{|r|r|r|r|r|r|}
\hline
$\lambda$& line  &\multicolumn{2}{c|}{F
$(erg\,cm^{-2}\,s^{-1})$}&\multicolumn{2}{c|}{EW (\AA)}\\
(\AA)  &     & This paper  &  KB95    &  This paper  &  KB95    \\
\hline
4659   &C~IV+He~II&$2.56\times 10^{-14}$&$1.48\times 10^{-14}$&$360\pm
25$&$330\pm20$\\
5290     &O~VI    &$9.50\times 10^{-16}$&$1.26\times 10^{-15}$&$ 13\pm
2$&$55       $\\
5411     &He~II+C~IV&$1.28\times 10^{-15}$&$5.5\times  10^{-16}$&$ 17\pm
2$&$30       $\\
5470     &C~IV    &$3.46\times 10^{-16}$&$4.2 \times 10^{-16}$&$  4\pm
1$&$20       $\\
5590     &O~V     &$4.04\times 10^{-15}$&$2.46\times 10^{-15}$&$ 62\pm
11$&$120      $\\
5801     &C~IV    &$2.10\times 10^{-14}$&$1.41\times 10^{-14}$&$420\pm
97$&$690\pm40 $\\
\hline
\end{tabular}
\end{center}
\end{table*}

\begin{table*}
\caption{
Mean velocities for clumps A, B, and C and toward the WO star.
}
\begin{center}
\begin{tabular}{|r|r|r|r|r|r|}
\hline
$\lambda$& Line  &\multicolumn{4}{c|}{$\mbox{I}/\mbox{I}(H_\beta)=100$}\\
(\AA)   &     &   A  &  B    &   C  &  KB95    \\
\hline
4686 & He~II     & $ 31.1\pm 0.7$ & $ 28.3\pm 0.8$ & $ 12.2\pm 0.8$ & 23.1  \\
4711 & [Ar~IV]   & $  7.1\pm 0.7$ & $  7.9\pm 0.8$ & $  3.1\pm 0.8$ &  4.5  \\
4740 & [Ar~IV]   & $  6.3\pm 0.7$ & $  6.5\pm 0.7$ & $  5.8\pm 0.8$ &  5.62 \\
4861 & $H_\beta$ & $100.0       $ & $100.0       $ & $100.0       $ & 100.0 \\
4959 & [O~III]   & $174  \pm 5  $ & $168  \pm 8  $ & $164.0\pm 2  $ & 188   \\
5007 & [O~III]   & $519  \pm 13 $ & $511  \pm 20 $ & $489  \pm 6  $ & --    \\
5876 & He~I      & $  4.1\pm 0.3$ & $  4.6\pm 0.7$ & $  5.4\pm 0.8$ & 7.1   \\
6542 & [N~II]    & $  1.4\pm 0.2$ & $  1.8\pm 0.3$ & $  2.3\pm 0.3$ & --    \\
6563 & $H\alpha$& $273  \pm 4.0$ & $283  \pm 10 $ & $273  \pm 5  $ & 276   \\
6583 & [N~II]    & $  2.4\pm 0.2$ & $  2.8\pm 0.4$ & $  4.6\pm 0.3$ & 6.4   \\
6717 & [S~II]    & $  8.9\pm 0.3$ & $ 10.2\pm 0.2$ & $ 15.2\pm 0.9$ & 8.73  \\
6731 & [S~II]    & $  6.9\pm 0.3$ & $  7.6\pm 0.2$ & $ 10.4\pm 0.8$ & 5.78  \\
\hline
\multicolumn{2}{|c|}{$I({H_\beta})\, erg \,cm^{-2}\, s^{-1}$}&
                   $1.5\times10^{-14}$ &   $1.2\times10^{-14}$ &
$7.7\times10^{-15}$ & \\
\hline
\end{tabular}
\end{center}
\end{table*}

The resulting mean clump velocities are consistent with the mean
velocity of the nebula S3 as a whole derived from our Fabry-Perot
interferometric   measurements   ($V_{HEL}  =  -216  \pm  1\km$,
Paper~I, taking
into account possible systematic absolute calibration errors of
$10-20\km$ for the interferometric observations), and are in full
agreement with the measurements of Tomita et al. (1993), based on a
single spectrogram of S3: $V_{HEL} = -235$ to $-230\km$. We show
below (see Section 3) that the velocities of the densest H~I clouds in
the vicinity of S3 are $\approx -230\km$ or differ from this value
by less than $10\km$.

\begin{figure*}
\psfig{figure=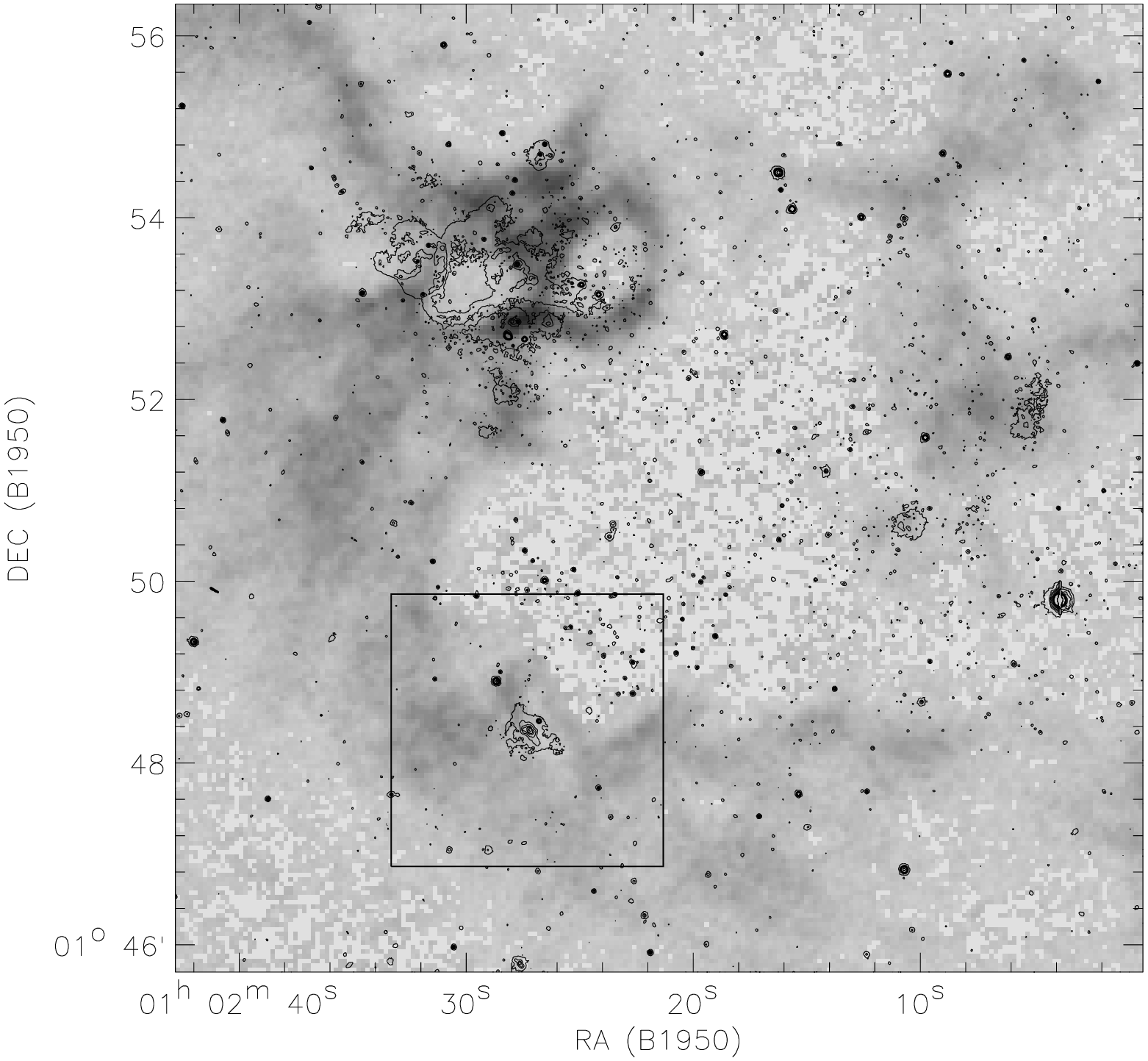,width=16 cm}
\caption{
Eastern sector of the galaxy IC 1613: brightness
distribution in the 21-cm line from the VLA data (grey scale) and
H$\alpha$ isophotes (based on the observations made with the 4-m KPNO
telescope). An enlarged-scale version of the framed region surrounding
the nebula S3 and WO star is shown in Fig. 4.
}
\end{figure*}
Two cone-shaped regions where the velocities are systematically
different from the mean are immediately apparent in Fig. 2.
Our [O~III] line measurements enable identification of the region
where the velocities differ from the mean value by $40-50\km$.
Both deviations from the mean velocity agree with the velocity variations
in the central part of the bright nebula S3 presented in Paper~I, based
on interferometric H$\alpha$ measurements. Note that the cone-shaped
regions originate at the position of the star, a pattern that, generally
speaking, might indicate an asymmetric stellar wind. However, firm
conclusions require further observational evidence, since such velocity
variations could also be a consequence of the inhomogeneous structure
of the interstellar medium in the near environment of the star.

\begin{table*}
\caption{
Mean velocities of clumps A, B, C, and toward the WO star.
}
\begin{center}
\begin{tabular}{|l|l|l|l|l|}
\hline
Lines &\multicolumn{4}{c|}{ $V_{HEL}, \km$}\\
      &    A        &      B       & C           & WO        \\
\hline
HeII 4686  & $-213\pm14$ & $-227\pm22 $ & $-186\pm30$ & $-214\pm21$ \\
$H_\beta$  & $-250\pm10$ & $-242\pm13 $ & $-252\pm7.$ & $-231\pm18$ \\
$$[OIII]4959& $-247\pm11$ & $-241\pm13 $ & $-257\pm6.$ & $-230\pm18$ \\
$$[OIII]5007& $-241\pm8.$ & $-238\pm17 $ & $-253\pm10$ & $-225\pm18$ \\
$H\alpha$ & $-246\pm6.$ & $-248\pm5. $ & $-242\pm4.$ & $-251\pm6.$ \\
$$[SII]  6717& $-247\pm23$ & $-225\pm13 $ & $-224\pm11$ & $-253\pm26$ \\
\hline
$<V_{HEL}>$&$-241\pm6 $ & $-237\pm4  $ & $-236\pm11$ & $-234\pm6 $ \\
\hline
\end{tabular}
\end{center}
\end{table*}

\bigskip

\section{ Radio observations in the 21-cm line}

\subsection{ Observations and Data Reduction}

We constructed a high angular resolution VLA H~I map covering an
extended region of the galaxy IC 1613, including the WO star and the
associated bipolar shell (this was part of a large project to
investigate  the  neutral-gas  structure and kinematics in IC 1613
by Wilcots et al., in press).
The maps shown below were constructed from combined B, C, and D-array
VLA observations with a velocity resolution of $2.57\km$. The data
were Hanning smoothed, subjected to the usual calibration procedure,
and made into maps using the AIPS package. The results are
represented in the form of a data cube with angular resolution
$7.^{\prime\prime}4\times7.^{\prime\prime}0$, which corresponds to a
linear resolution of $\sim$23 pc.

Figure 3 shows the H~I distribution for the eastern sector of the
galaxy constructed using these data. A well-defined giant cavity
(supercavity) surrounding a large fraction of the stellar population
of IC 1613 is immediately apparent. This giant, low surface brightness
region of 21-cm emission is also visible on the map of Lake \&
Skillman (1989). Earlier Paper~I, we suggested that this region was deficient
in neutral gas because it had been swept out, implying that the
supercavity should be surrounded by a dense, wind-blown shell. Figure 3
shows that such a shell or ring of neutral gas does, indeed, exist, and
that the WO star resides in a dense gas layer near its inner edge.

The angular size of the supercavity is about $5^{\prime}$, corresponding
to a linear size of 1~kpc. The characteristic thickness of the shell is
$\Delta R \simeq 1-2^{\prime}$, or 200--350~pc.

A sketch of the bipolar ionized shell surrounding the WO star
and its relative position with respect to the H~I supercavity are
shown on an enlarged scale in Fig. 4, which presents the southeastern
sector of the H~I ring. This figure shows that the core of the ionized
bipolar shell (previously known as S3) and its bright southeastern
component are located in a dense region of the ring, whereas the extended,
faint northwestern component is in the low-density medium inside the
supercavity.

\begin{figure*}
\centerline{\psfig{figure=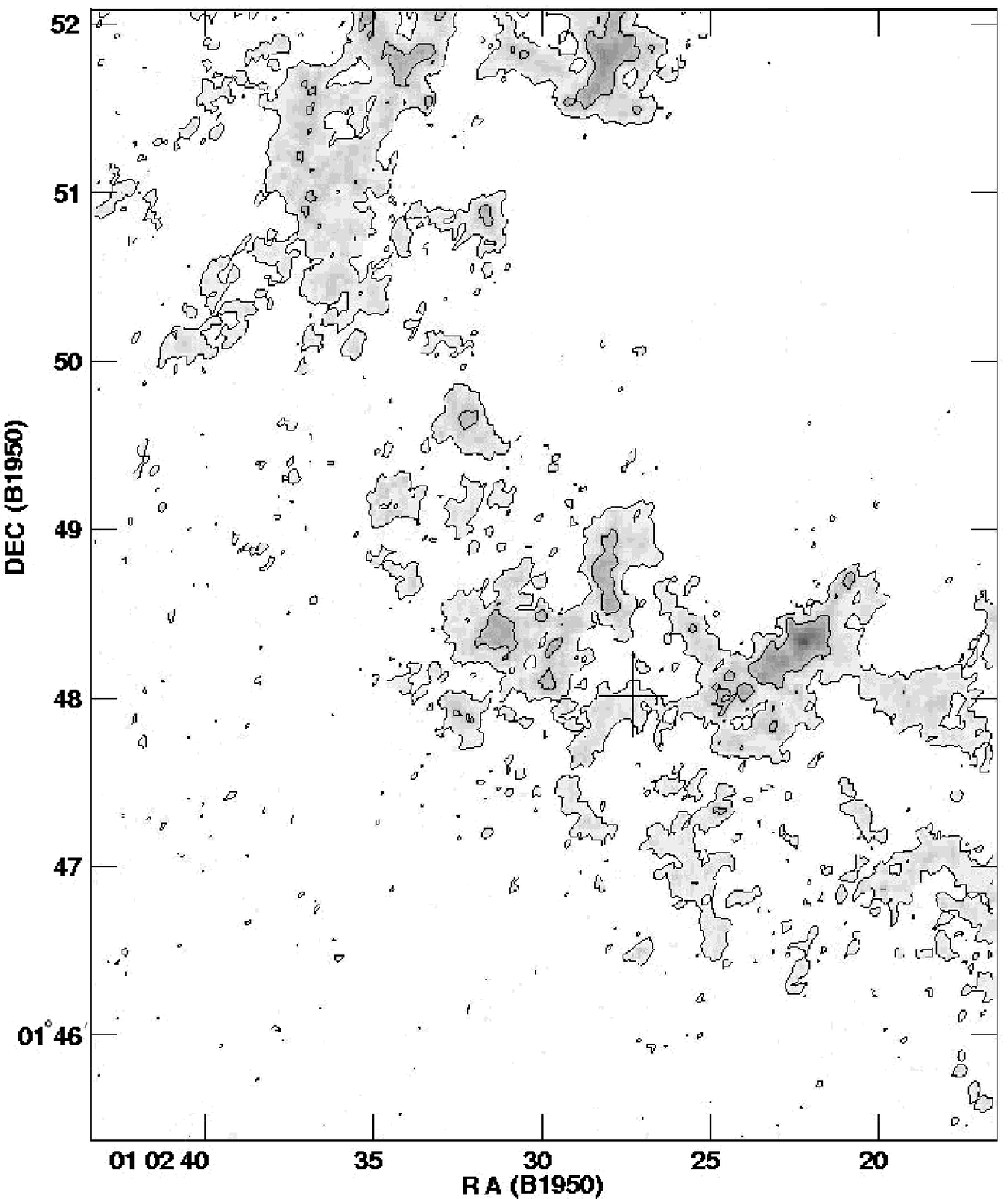,width=9 cm}}
\centerline{\psfig{figure=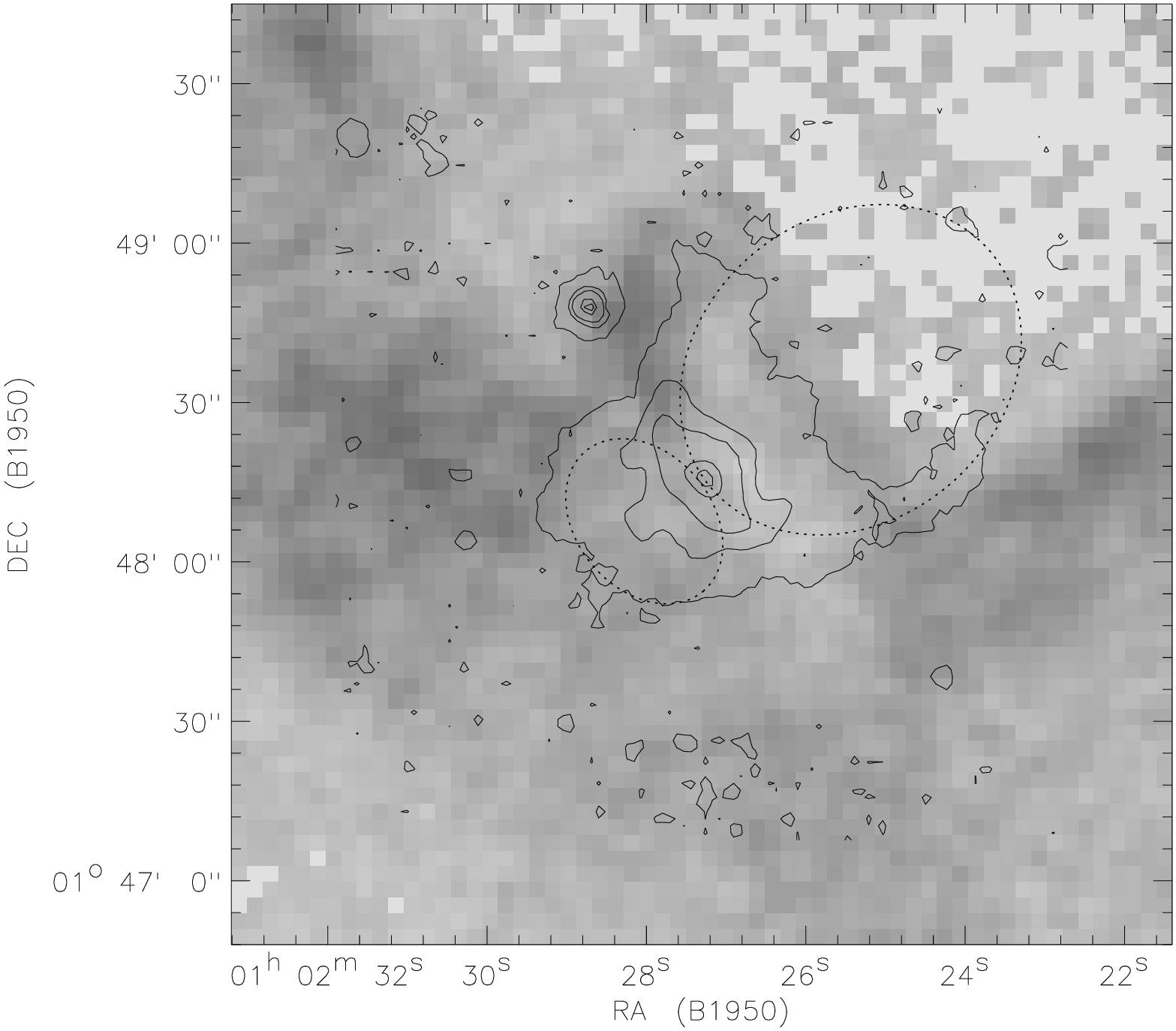,width=9 cm}}
\caption{
{\it top} -- Southeastern sector of the dense H~I ring. The contours
correspond to H~ I column densities
of N(H~I) = $(2, 6, 18)\times 10^{20}$~cm$^{-2}$. The cross indicates the
position of the WO star. {\it bottom} -- Sketch of the bipolar ionized shell
associated with the WO star (isophotes adopted from Paper~I); the H$\alpha$
isophotes correspond to brightness levels of 0.028, 0.2, 0.4, 1.0, and
$2.0\times 10^{-15}$~erg~cm$^{-2}$~s$^{-1}$~arcsec$^{-2}$. The H~I
distribution is superimposed as a grey-scale image (including the
framed region in Fig. 3).
}
\end{figure*}

Figure 4 clearly shows a local decrease in the 21-cm brightness,
filled by the ionized gas of the core and southeastern component.
We can also see two prominent, extended H~I features just on the outside
of the two sides of the northwestern component of the ionized shell,
which reproduce the shape of the shell. This structure probably provides
evidence for a local neutral shell surrounding the bipolar shell.

The parameters of the most prominent H~I clouds forming the local
neutral shell surrounding the ionized bipolar feature are summarized
in Table~5, which presents the central coordinates, total mass, and
maximum column density of each cloud.

\begin{table*}
\caption{
Parameters of HI clouds in the vicinity of the bipolar
shell.
}
\begin{center}
\begin{tabular}{|c|l|l|c|c|}
\hline
Clouds & RA(B1950) & Dec(B1950) & N(HI), $cm^{-2}$ & $M,\,M_{\odot}$    \\
\hline
 a     & 1 02 28    & 1 47 55 & $5\times10^{20}$ & $\sim4\times10^{4}$ \\
 b     & 1 02 31.5  & 1 48 20 & $8\times10^{20}$ & $\sim3\times10^{4}$ \\
 c     & 1 02 21-24 & 1 48 08 & $2\times10^{21}$ & $    1\times10^{5}$ \\
 d     & 1 02 26    & 1 48 20 & $7\times10^{20}$ & $    2\times10^{4}$ \\
 e     & 1 02 28    & 1 48 30-60& $9\times10^{20}$ & $   6\times10^{4}$ \\
\hline
\end{tabular}
\end{center}
\end{table*}

\subsection{ Neutral Gas Velocities}

In order to reveal the possible effect of the expansion of the local
neutral shell, we constructed the radial-velocity distribution within
a right-ascension band passing through the WO star (Fig. 5). The
band is $200^{\prime\prime}$ wide and therefore includes all clouds
making up the local shell. This distribution clearly demonstrates
velocity variations with distance from the star that are typical
of an expanding shell (the so-called `velocity ellipse') at right
ascensions from $01^h 02^m 28^s$ to $01^h 02^m 31^s$. This region
corresponds to the part of the local
H~I shell that is east of the WO star; the characteristic radius
of this `half-shell' is 100--140 pc. The velocity difference between
the two sides of the neutral shell is $\sim20\km$, suggesting an expansion
velocity of $\sim10\km$. The velocities of the densest `unaccelerated'
H~I clouds in a large region surrounding the WO star are V$_{Hel} =
-230\pm 5\km$.

\begin{figure}
\psfig{figure=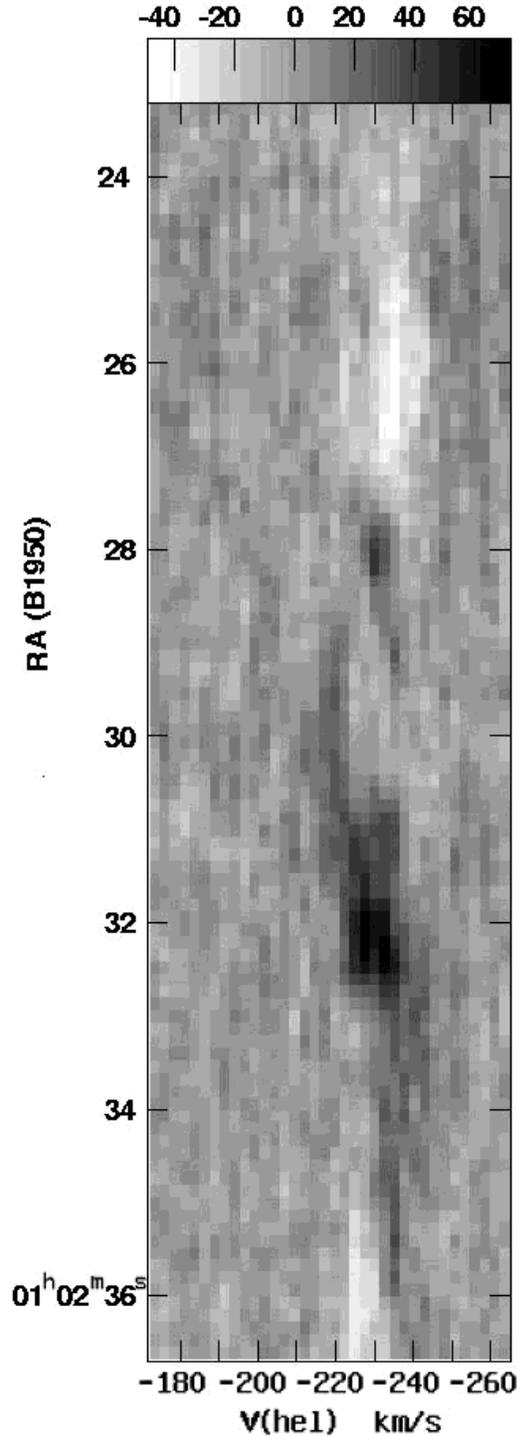,width=7 cm}
\caption{
H~I radial-velocity variations in a right-ascension band
from the VLA observations (see text). The fluxes are in mJy/beam.
}
\end{figure}

\section{Discussion}

     The  results of spectroscopic observations of the WO star and
the  central region of the nebula S3 reported in this paper are in
full  agreement with earlier observations, within the errors, when
the  different  slit locations are taken into account. The spectra
of  the  compact  clumps  identified  in  this paper do not differ
significantly  from the those of the nebula S3 (see Garnett et al.
(1991),  KB95 and references therein). The
relative  line  intensities  listed in Table 3 are suggestive of a
typical  H~II  region  ionized  by  a  star  with  high  effective
temperature.  The  observed differences between the spectra of the
three  clumps  could  be  due  to different excitation conditions:
distance  from  the  exciting  star and electron density. No shock
signatures  were  found either in the emission spectra (the [N~II]
and  [S~II]  lines  are  weak)  or  in  the mean velocities of the
bright  clumps.  The  ionized  gas velocities of the clumps in the
central  core  of  the  nebula agree with velocity measurements of
neutral gas clouds in the vicinity of the bipolar shell.

\subsection{ Electron density and He abundance in the nebula core.}

  The  relative  intensities of the [S~II]  doublet
can  be  used to estimate the electron density in the three bright
clumps  identified  in  this  paper.  We  determined  the electron
densities  in  clumps  A,  B,  and  C  using formulas adopted from
De Robertis et al.(1987),  assuming  that the electron temperature of the
nebula   is   $T_e  =  17  000$~K  (as  determined  in  KB95  from
[O~III]~4959/4363\AA~ line  intensities). The resulting values are
listed in Table~6.

We derived the He abundance in the nebula from He~I and He~II
line intensities of individual clumps using the diagnostic relations
and recombination coefficients for hydrogen and helium from Kholtygin
(1997),. We determined the relative helium abundances N(He$^+$)/N(H) and
N(He$^{++}$)/N(H) and the total helium abundances N(He)/N(H) (in
number) of the clumps. The results are given in Table 6, which also
lists data adopted from KB95. We can see that our abundances coincide
with those of KB95, and are in good agreement with the estimate of
Davidson\& Kinman (1982):  N(He)/N(H) = $0.073 \pm 0.009$. (In both Davidson \& Kinman
(1982)  and KB95, the He
abundances were obtained from the spectrum of the entire bright
core of the nebula without subdividing it into individual clumps).

\subsection{ Velocity of the stellar wind of the WO star. }

One of the major parameters determining the structure and
kinematics of the interstellar medium at large distances from WR
stars is the stellar-wind velocity. Kingsburgh et al.(1995) have
shown (see also KB95) that the halfwidth of the C~IV~5801\AA~line
at zero intensity is a good estimator of the terminal wind velocities
of WR stars. The C~IV line halfwidth at zero intensity measured from
our spectra for the WO star yields a terminal wind velocity of
$V_w = 2910\km$, which is virtually coincident with the value
$2850\km$ obtained in KB95, given that the accuracy of our measurements
is determined by the width of the spectral channel ($\sim 130\km$).
We thus confirm the conclusion of KB95 that
the terminal wind velocity of the WO star in IC 1613 is the lowest
among the six objects of this class currently known in the Local-Group
galaxies.

\subsection{ Large-scale Structure of the Interstellar Medium. }

In principle, the giant H~I ring shown in Fig. 3 could be the
projection onto the plane of the sky of a shell or torus. A torus
could be the result of the wind-blown supercavity breaking out of the
gaseous disk of the galaxy. In both cases, the neutral gas ring
surrounding the supercavity displays a characteristic filamentary
structure. The filamentary system making up the H~I ring bifurcates
at its southeastern end. A system of arc-like filaments with characteristic
lengths of $1-1.5'$, or 200--300 pc, is also apparent. This filamentary
pattern undoubtedly testifies to the action of dynamically active
processes that are responsible for the formation of the supercavity
and the surrounding dense ring or shell. This large-scale structure
apparently formed as the result of a burst of star formation in the
region. However, determining the nature of this burst is a separate
problem, which lies beyond the scope of this paper. Of interest to us
here is the very existence of such a dense shell or torus surrounding
the supercavity, since we consider this large-scale mass distribution
to be responsible for the unique bipolar structure of the extended
ionized shell associated with the WO star.

\begin{table}[t]
\caption{
Electron density and He abundances of clumps A, B, and C.
}
\begin{tabular}{|l|l|l|l|l|}
\hline
                    &    A  &  B     & C        & KB95     \\
\hline
$ n_{e}\, cm^{-3}$  & 170   & 80     & $\le 30$ & 100      \\
\hline
 N(He$^{+}$)/N(H)   & 0.051 & 0.057  & 0.067    & 0.057    \\
 N(He$^{++}$)/N(H)  & 0.028 & 0.025  & 0.011    & 0.025    \\
 N(He)/N(H)         & 0.079 & 0.082  & 0.078    & 0.082    \\
\hline
\end{tabular}
\end{table}
Lozinskaya (1997)  suggested that the bipolar structure of the
ionized shell surrounding the WO star might be the result of a strong
stellar wind breaking out of the dense gas layer. Earlier Paper~I, we
associated this dense layer with a hypothetical wind-blown shell,
which we suggested should surround the giant H~I-deficient region
in IC 1613 seen on the map of Lake \& Skillman(1989). However, only
in this paper have we finally been able to provide some evidence
supporting this hypothesis. The following factors form our principal
arguments.

\begin{enumerate}
\item One piece of evidence is the relative orientation of the bipolar
ionized shell with respect to the H~I supercavity. It is clear from
Fig. 4 that the bright core of the nebula is extended along the boundary
of the dense H~I ring; the smaller southeastern component of the ionized
bipolar shell lies inside a dense neutral gas layer, and the extended
northwestern component lies in the low-density medium inside the
supercavity.
\item Afanas'ev et al. (2000) showed that the southeastern component
is brighter than the northwestern component, and also has a lower
expansion velocity.
\item The local cavity and extended H~I features whose discovery is
reported in this paper testify to the presence of a neutral shell
surrounding the ionized bipolar shell, providing direct evidence for a
physical association of the WO star and the ionized bipolar shell with
the supercavity and dense H~I ring (the only alternative being a
chance projection).
\end{enumerate}

The gas density in the clouds making up the local H~I shell
surrounding the bright central and southeastern components of the
ionized nebula can be derived from our maximum column density
and total mass estimates, assuming that the clouds have the same
sizes in the line-of-sight and transverse directions. The clouds that
make up the local neutral H~I shell have irregular structures. Clouds
a, b, and e in Table 5, which make up the expanding part of the H~I
half-shell, have lengths in the plane of the sky ranging from 40 to
120~pc. For a column density of $(5-9)\times 10^{20}$~cm$^{-2}$, this
yields a characteristic density of $n_{\mbox{H~I}}\simeq 1.5 - 7.5$~cm$^{-3}$.

For comparison, we present estimates of the average density of ionized
gas in this nebula: Davidson \& Kinman (1982) found $n_e = 8.5$~cm$^{-3}$
based on the H$\beta$ emission of the central $6^{\prime\prime}$ of the
nebula S3; Kenicutt (1984) reports $n_e \simeq 1$~cm$^{-3}$ based on a study
of the integrated H$\alpha$ emission of the bright nebula; and Goss \&
 Lozinskaya (1995) found $n_e = 3.5$~cm$^{-3}$ from their analysis of
its thermal radio emission.

     If   the  H~I  clouds  surrounding  the  bright  central  and
southeastern  components of the bipolar ionized shell are, indeed,
parts  of  a  shell  of neutral gas blown out by the stellar wind,
the  radius  and  expansion velocity of the expanding section east
of   the   WO  star  should  be  $100-140$~pc  and  about  $10~\km$,
respectively  (see  Section  3.2).  In the standard model (Castro et
al.,1975;  Weaver  et  al.,1975), we find the kinematic age of the
wind-blown  neutral  shell  to be $t = (6-8)\times 10^6$~yr. (Note
that  this  estimate depends neither on the stellar wind power nor
on  the  density  of  the surrounding gas). This age substantially
exceeds  the  duration  of  the final WO stage, and corresponds to
the   main-sequence  lifetime  of  a  massive  star  ($M\ge  30-40
M_{\odot}$).  We  thus  conclude  that  the  local  neutral  shell
surrounding  the bright southeastern and central components of the
ionized  nebula  could  have  been produced by the stellar wind of
the  WR  progenitor  when  it was still a main-sequence star. Note
that,  earlier  Paper~I,  we concluded that the ionized bipolar shell
was  blown  out  by the successive action of the stellar wind from
the central star, first at the WR and then at the WO stage.

Our estimates indicate that the terminal velocity of the stellar wind
of the WO star reaches $3000\km$ (see Section 4.2); a mass-loss rate
of $2.9\times 10^{-5}M_{\odot}$/yr was derived in KB95 based on the relation
between the He~II line luminosities and mass-loss rates for two WO stars.
Accordingly, the power of the stellar wind of the WO star in IC 1613 is
$L_w \simeq 10^{38}$~erg/s. In all likelihood, the wind power was a factor of
two to three lower during the earlier WR stage (WR stars have typically
terminal velocities of $2000\km$), but the duration of the WR stage
($5\times 10^5$~yr) is at least an order of magnitude longer than that of
the WO stage. The power of the stellar wind of the main-sequence
progenitor could have been two orders of magnitude lower than at
the final Wolf-Rayet stage, however, the main-sequence lifetime was at
least an order of magnitude longer than the Wolf-Rayet stage.

The formation of the wind-blown shell is determined by the
standard relation:
     $$  R  = 28 (L_{36}/n_{o})^{1/5} t_{6}^{3/5} $$, where $L_{36}$
is  the  power  of  the  stellar wind in units of $10^{36}$~erg/s,
$n_o$  the  density  of the surrounding gas, $R$ the radius in pc,
and  $t_6$  the  time  in  units of $10^6$~yr (Castro et al.,1975;
Weaver  et  al.,1975).  We  can  easily  see  that,  although  the
mechanical  energy  inputs from the massive star during the WR and
WO  stages  are  comparable,  the stellar wind produced during the
main-sequence  stage  is  capable  of creating a larger shell than
the  powerful  but  short-lived  stellar  wind at the final stage.
Therefore,  this  latter  wind  serves  as an additional source of
mechanical energy in an already formed, slowly expanding shell.

Our results for the two extended arc-like H~I clouds surrounding
the northwestern component of the ionized shell are much less
conclusive. On the one hand, these formations could be the result
of the stellar wind of the WO star and its progenitor breaking into
the tenuous gas inside the supercavity. Our measurements, indeed,
indicate that the H~I surface density inside the supercavity is at least
an order of magnitude lower than in the clouds listed in Table 5.
The thickness of the emitting gas layer can be taken to be twice that
of the supershell in the plane of the sky; i.e., about $3-4^{\prime}$
(600--700~pc),
according to Fig. 3. Accordingly, we find the mean gas density in the
supercavity to be $\simeq 0.04-0.05$~cm$^{-3}$. If the observed ring-shaped
H~I distribution is the projection of a thick torus rather than a
shell, the size of the emission region should be even larger, so that
the above estimate should be considered an upper limit to the gas density
inside the supercavity. The same standard relation yields a radius
of $R \simeq 180$~pc for a shell blown out in the tenuous medium of the
estimated density over the time $t = 7 \times 10^6$~yr found above. This
agrees with the observed size of the northwestern part of the local
shell outlind by two extended arc-like H~I clouds in Fig. 4.

On the other
hand, the formation of such an arc-like structure outlined by these
extended H~I clouds could be the result of the collective stellar winds
(and possible supernova explosions) from association No.~9 of the list
of Hodge (1978), which is adjacent to and inside the supercavity. (This
association is the largest in size and the richest in the galaxy). In
this case, the stellar wind and ionizing radiation of the WO star act
inside this already formed arclike feature.

We plan a detailed analysis of the kinematics of the ionized and
neutral gas components in this region, and a comparison with the
stellar population. This study should lead to unambiguous conclusions
about the nature of the arclike formation and the possible genetic relation
between the WO star and the association.

\section{ Conclusions }

We have observed the central part of the nebula S3 surrounding the
WO star (field spectroscopy with the 6-m telescope) and an extended
surrounding region more than one kpc in size (21-cm VLA observations). We
have mapped the brightness of the central nebula in the main
spectral lines and constructed the radial-velocity distribution based
on H$\beta$ and [O~III]~5007\AA~ line measurements.

We identified three compact clumps, and determined their integrated
spectra, mean velocities, and electron densities. Our spectroscopic
data for both the WO star and the central region of S3 are in agreement
with previous observations (see Garnett et al.(1991), KB95 and references
therein). The
relative line intensities in the spectrum of the nebula are indicative
of pure recombination emission from a highly excitated H~II region.

We analyzed the structure and kinematics of the neutral gas in an
extended region around IC 1613 that includes the WO star and the
associated ionized bipolar shell. A giant, dense shell or toroidal
formation of neutral gas about 1.5 kpc in size surrounds the H~I
supercavity. This giant structure was very likely produced by a burst
of star formation in this region.

We also discovered for the first time a local deficiency in the
brightness of the 21-cm line and two extended H~I features, providing
evidence that the extended ionized bipolar shell associated with the
WO star (Lozinskaya, 1997 and  Paper~I)
 is, in turn, surrounded by a local shell of neutral
gas. The mass and expansion velocity of this shell indicate that it
was blown out by the stellar wind of the WO progenitor when it was
still a main-sequence star.

Our new observations support a scenario in which the extended bipolar
structure was produced by the stellar wind of the WO star residing in
the dense `wall' of the supercavity. The arclike structure of the
extended H~I clouds just outside the faint northwestern component of
the ionized shell might have been shaped by the collective stellar
wind (and supernova explosions?) of Hodge's (1978) association no. 9. In
this case, the wind and ionizing radiation of the WO star act inside this
previously formed arclike structure.

\acknowledgements{
\footnotesize
This work was supported by the Russian Foundation for Basic Reearch
(project code 98-02-16032) and the Astronomy State Science and Technology
Program (project no. 1.3.1.2). The National Radio Astronomy Observatory
is a facility of the National Science Foundation operated under
cooperative agreement with Associated Universities, Inc. We are grateful
to E. Blanton for sharing the H$\alpha$ line image of IC 1613 taken at
Kitt Peak National Observatory, E. E. Gerasimenko for supporting the
observations at the 6-m telescope of the Special Astrophysical Observatory
of the Russian Academy of Sciences, and the Program Committee of the 6-m
telescope for providing the observing time for this project.
}

\end{document}